\documentclass[12pt]{article}
 \usepackage{epsfig}
 \def\be{\begin{equation}}
 \def\ee{\end{equation}}
 \def\bea{\begin{eqnarray}}
 \def\eea{\end{eqnarray}}
 \usepackage{graphicx}

 \catcode`\@=11
 \def\lsim{\mathrel{\mathpalette\@versim<}}
 \def\gsim{\mathrel{\mathpalette\@versim>}}
 \def\@versim#1#2{\vcenter{\offinterlineskip
 \ialign{$\m@th#1\hfil##\hfil$\crcr#2\crcr\sim\crcr } }}
 \catcode`\@=12

 \catcode`@=12
\begin{document}
 \thispagestyle{empty}
 \begin{flushright}
 UCRHEP-T627\\
 Nov 2023\
 \end{flushright}
 \vspace{0.6in}
 \begin{center}
 {\LARGE \bf Universal Symmetry and its Soft Breaking in\\ 
Renormalizable Supersymmetric Field Theory\\}
 \vspace{1.5in}
 {\bf Ernest Ma\\}
 \vspace{0.1in}
{\sl Department of Physics and Astronomy,\\ 
University of California, Riverside, California 92521, USA\\}
 \vspace{1.2in}

\end{center}

\begin{abstract}\
It is pointed out that every renomalizable supersymmetric field theory has a 
symmetry which is hidden in plain sight, but is usually broken by soft terms 
which obey supersymmetry.  On the other hand, the terms which break 
supersymmetry softly may or may not break this symmetry.  Implications for 
the minimal supersymmetric standard model (MSSM) are discussed.
\end{abstract}

\newpage
\baselineskip 24pt

\noindent \underline{\it Introduction}~:~ 
It has been pointed out recently~\cite{m23} that every renormalizable 
field theory has a $Z_4$ symmetry under which vector gauge bosons, scalars, 
left-handed fermions, and right-handed fermions transform respectively 
as $(1,-1,i,-i)$, 
broken only by soft explicit dimension-three terms or spontaneously. 
Here it is further pointed out that an additional softly broken symmetry 
exists if the field theory comes from supersymmetry, as does the 
minimal supersymmetric standard model (MSSM).

If spacetime is extended to include superspace, fermions and bosons may be 
combined to form superfields with supersymmetry.  There are two basic 
superfields.  A matter superfield $\hat{\phi}$ consists of a complex 
scalar $\phi$ and a chiral fermion $\psi$ (conventionally chosen as 
left-handed). 
A gauge superfield $\hat{A}_\mu$ consists of a massless vector gauge 
field $A_\mu$ and its Majorana fermion counterpart $\lambda$ (gaugino). 

\noindent \underline{\it Observation}~:~
A renormalizable supersymmetric field theory is obtained if the 
superpotential $\hat{W}$ of the theory is the sum of bilinear and trilinear 
products of matter superfields, i.e. of the form
\begin{equation}
\hat{W} = \mu_{ij} \hat{\phi}_i \hat{\phi}_j + f_{ijk} \hat{\phi}_i 
\hat{\phi}_j \hat{\phi}_k.
\end{equation} 
It is thus obvious that a discrete symmetry $Z_3$ exists under which 
$\hat{A}_\mu \sim 1$ and $\hat{\phi} \sim \omega$, 
with $\omega^3 = 1$.  Of course, this symmetry is broken completely by 
a possible bilinear term, but only softly, so its associated mass parameter 
may be chosen to be small, as advocated in Ref.~\cite{m23} for the 
$Z_4 \to Z_2$ symmetry breaking which applies to all renormalizable field 
theories.

\noindent \underline{\it Insight}~:~ 
In supersymmetry, all scalar fields are complex.  Hence the $Z_3$ 
transformation is possible, which also applies to the chiral fermion in 
the matter superfield.  However, the Majorana fermion (gaugino) in the gauge 
superfield is neutral under $Z_3$.  Thus the gaugino mass is allowed by 
$Z_3$ but breaks supersymmetry as well as the $Z_4$ symmetry of 
Ref.~\cite{m23}.  There are now several kinds of soft terms, depending 
on whether or not supersymmetry, $Z_3$, or $Z_4$ are broken.  They will 
be discussed accordingly in the context of the MSSM.

\noindent \underline{\it Higgs potential in the MSSM}~:~
It is well-known that there are two Higgs doublets in the MSSM, 
$\Phi_1 = (\phi_1^0,\phi_1^-)$ and $\Phi_2 = (\phi_2^+,\phi_2^0)$, 
with $v_1 = \langle \phi_1^0 \rangle$ giving mass to the $d$ quarks 
and charged leptons, and $v_2 = \langle \phi_2^0 \rangle$ giving mass 
to the $u$ quarks.  The parameter $\tan \beta = v_2/v_1$ is unknown, 
but is often taken to be large because $m_t$ is much greater than 
$m_b$ and $m_\tau$.  The superpotential of the MSSM is of the form 
\begin{equation}
\hat{W} = \mu \hat{\Phi}_1 \hat{\Phi}_2 + f_d \hat{\Phi}_1 \hat{Q} \hat{d}^c + 
f_l \hat{\Phi}_1 \hat{L} \hat{l}^c + f_u \hat{\Phi}_2 \hat{Q} \hat{u}^c,
\end{equation}
where $Q = (u,d)$ and $L = (\nu,l)$ are the usual quark and lepton doublets.
Under $Z_3$, each matter superfield transforms as $\omega$, hence the 
$\mu$ term breaks it softly.  Its 
contribution to the Higgs potential $V$ is just 
$|\mu|^2 (\Phi_1^\dagger \Phi_1 + \Phi_2^\dagger \Phi_2)$.

The quartic couplings of $V$ come from the gauge interactions.  For 
easy comparison to the usual two-Higgs-doublet model, $\Phi_1$ is 
redefined to denote $(\phi_1^+,-\bar{\phi}_1^0)$.  In that case,
\begin{eqnarray}
V &=& m_1^2 \Phi_1^\dagger \Phi_1 + m_2^2 \Phi_2^\dagger \Phi_2 + m^2_{12} 
(\Phi_1^\dagger \Phi_2 + \Phi_2^\dagger \Phi_1) + 
{1 \over 8} (g_1^2 + g_2^2) [(\Phi_1^\dagger \Phi_1)^2 + 
(\Phi_2^\dagger \Phi_2)^2] \nonumber \\ &-& {1 \over 4} (g_1^2-g_2^2) 
(\Phi_1^\dagger \Phi_1)(\Phi_2^\dagger \Phi_2) - {1 \over 2} g_2^2 
(\Phi_1^\dagger \Phi_2)(\Phi_2^\dagger \Phi_1),
\end{eqnarray}
where $g_1$ and $g_2$ are the $U(1)$ and $SU(2)$ gauge couplings of the SM. 
The $|\mu|^2$ term from $\hat{W}$ has been absorbed into 
$m_{1,2}^2$ with $m_1^2-m_2^2$ and $m_{12}^2$ as the soft supersymmetry 
breaking terms.  Note that the only term which breaks $Z_3$ is $m_{12}^2$. 
The mass spectrum of the five physical scalar particles of $V$ is 
well-known, i.e.
\begin{equation}
m_A^2 = m_1^2 + m_2^2 = m_{12}^2 (\tan \beta + \cot \beta), ~~~ 
m^2_{H^\pm} = M_W^2 + m_A^2, ~~~ 
\end{equation}
and the mass-squared mass matrix spanning the SM Higgs 
$h = \sqrt{2} Re(\cos \beta \phi_1^0 + \sin \beta \phi_2^0)$ 
and its orthogonal counterpart 
$H = \sqrt{2} Re(\sin \beta \phi_1^0 - \cos \beta \phi_2^0)$ is given by
\begin{equation}
{\cal M}^2_{hH} = \pmatrix{M_Z^2 \cos^2 2 \beta & -M_Z^2 \sin 2 \beta \cos 2 
\beta \cr -M_Z^2 \sin 2 \beta \cos 2 \beta & M_A^2 + M_Z^2 \sin^2 2 \beta}.
\end{equation}
Since $m^2_{12}$ is the only term which breaks $Z_3$, it should be small 
compared to $m_A^2$.  Hence $\sin 2 \beta = 2 m^2_{12}/m_A^2 << 1$.  Thus
$\cos^2 2 \beta \simeq 1$ and $m_h$ is almost equal to $m_Z$.  However, 
to agree with the observed $m_h = 125$ GeV, radiative corrections 
(especially from the top quark couplings) must be considered~\cite{dr16}. 
In any case, the observation $m^2_{12} << m_A^2$ is consistent with this 
scenario.

\noindent \underline{\it Supersymmetry breaking terms}~:~
Whereas the standard model (SM) particles obtain their masses from 
$v_{1,2}$, their supersymmetric partners are much heavier, with 
contributions from soft supersymmetry breaking terms.  The observation 
of this paper is that these terms may be classified according to how 
they transform under $Z_3$ and $Z_4$, as shown in Table~1. 
\begin{table}[tbh]
\centering
\begin{tabular}{|c|c|c|}
\hline
mass parameter & $Z_3$ & $Z_4$ \\ 
\hline
$m^2_{12}$, $B$ & $\omega^2$ & 1 \\ 
\hline
$M_{1,2,3}$, $A$  & 1 & $-1$ \\ 
\hline
$\tilde{m}^2$ & 1 & 1 \\
\hline
\end{tabular}
\caption{Soft supersymmetry breaking parameters under $Z_3$ and $Z_4$.}
\end{table}
The superpartner scalar masses $\tilde{m}$ are allowed by both $Z_3$ and 
$Z_4$, whereas the gaugino masses $M_{1,2,3}$ respect $Z_3$ but break 
$Z_4$.  The former should then be chosen much larger than the latter 
from the $Z_4 \to Z_2$ argument of Ref.~\cite{m23}.  Interestingly,  
this result is also the basic assumption of split supersymmetry~\cite{ad05}, 
proposed many years ago.  The $A$ terms have trilinear scalar interactions 
which also respect $Z_3$ but break $Z_4$.  As already pointed out, 
$m^2_{12}$ of Eq.~(3) (which is a $B$ term) breaks $Z_3$ but not $Z_4$. 
A natural hierarchy thus exists with $\tilde{m}$ much larger than the 
others. 

Higgsino masses depend on $\mu$ from Eq.~(2) which is the only mass 
parameter of the MSSM in the supersymmetric limit.  It breaks $Z_3$ 
but has no other mass scale to compare with, until after supersymmetry 
breaking.  Hence it should be small compared with $\tilde{m}$ which  
respects $Z_3$, and perhaps also smaller than the gaugino 
masses $M_{1,2,3}$ which also respect $Z_3$.  This 
suggests that the lightest neutralino and chargino are mostly 
Higgsinos and the former is a possible dark matter candidate,  
which is consistent with global fits~\cite{gambit17} to the 
present data.

\noindent \underline{\it Conclusion}~:~
If the MSSM is a viable extension of the SM, then the soft supersymmetry 
breaking terms may be classified according to the $Z_4$ of Ref.~\cite{m23} 
and the $Z_3$ proposed in this study.  A natural hierarchy of mass scales 
emerges which is consistent with present data.  However, since there is 
yet no experimental evidence for the existence of supersymmetry, these 
ideas are only waiting to be tested.

\noindent \underline{\it Acknowledgement}~:~
This work was supported in part by the U.~S.~Department of Energy Grant 
No. DE-SC0008541.  

\baselineskip 18pt
\bibliographystyle{unsrt}

\end{document}